\documentstyle[preprint,aps]{revtex}

\input epsf

\begin{document}
\draft
\title{Surface critical behaviour of the honeycomb
O($n$) loop model with mixed ordinary and special boundary
conditions}
\author{M T Batchelor and C M Yung}
\address{Department of Mathematics, School of Mathematical Sciences,
Australian National University, Canberra ACT 0200, Australia}
\date{May 1995}
\maketitle
\begin{abstract}
The O($n$) loop model on the honeycomb lattice with mixed
ordinary and special boundary conditions is solved exactly
by means of the Bethe ansatz. The calculation of the dominant
finite-size corrections to the eigenspectrum yields the mixed
boundary scaling index and the geometric
scaling dimensions describing the universal surface
critical behaviour. Exact results follow in the limit $n=0$ for
the polymer adsorption transition with a mixed adsorbing and free
boundary. These include the new configurational exponent
$\gamma_1=\mbox{\small $\frac{85}{64}$}$.
\end{abstract}
\pacs{\\{ANU preprint MRR 043-95\\}{\em To appear in J. Phys. A}}

The critical behaviour of semi-infinite $d$-dimensional
systems with free surfaces
can be very rich, with the possibility of ``special" or
multicritical surface behaviour when the surface couplings
are sufficiently enhanced \cite{bin,diehl}.
One fundamental model of relevance in this context is the
semi-infinite two-dimensional $n$-vector or O($n$) model \cite{s}.
In the limit $n=0$, which describes self-avoiding walks \cite{deg},
the special transition corresponds to the polymer adsorption
transition \cite{ekb}.

In this letter we derive the surface critical behaviour of the
O($n$) loop model on the honeycomb lattice with
``ordinary" (O) and ``special" (S) boundary
conditions on either side of a finite strip, i.e. a strip finite
in one direction and infinite in the other. Such mixed
boundary conditions are of current interest for
general $d$, being of relevance to the Casimir interaction between
two spherical particles in a fluid at the critical point (see, e.g.
\cite{er} and references therein).
In $d=2$, with boundary conditions $a$ and $b$ on opposite edges
of a strip of width $N$,
the free energy per site of a critical system scales as
(see \cite{be} and references therein)
\begin{equation}
f_N \simeq f_B + f_a N^{-1} + f_b N^{-1} +
\Delta_{a b} N^{-2}
\end{equation}
for large $N$. Here $f_B$ is the bulk free energy,
$f_a$ and $f_b$ are surface free energies and
$\Delta_{a b}$ is the universal Casimir amplitude.
This Casimir contribution
to the free energy has recently been
calculated for the O($n$) model with various mixed boundary conditions
via conformal-invariance methods \cite{be}.
Here we derive this quantity exactly for mixed O-S boundary conditions,
along with the surface scaling dimensions, from the Bethe Ansatz solution
of the corresponding O($n$) loop model \cite{dmns} on the honeycomb
lattice.

The partition sum of our O($n$) loop model is defined as
\begin{equation}
{\cal Z} = \sum x^{L} y_a^{L_a} y_b^{L_b} n^P
\end{equation}
where the sum is over all configurations of closed and nonintersecting
loops on the lattice depicted in figure 1.
Here $P$ is the total number of closed loops of fugacity $n$
in a given configuration. We take boundary condition $a$ ($b$) on the
left (right) edge of the strip. At $n=0$,
$x$ is the fugacity of a step in the bulk and $y_a$ ($y_b$)
is the fugacity of a step along the left (right) edge.
Thus $L$ is the number of steps in the
bulk and $L_a$ ($L_b$) is the number of steps along the
left (right) edge.

We have found that the equivalent 3-state vertex model can be solved
for a number of boundary conditions. The possible arrow configurations
and their corresponding Boltzmann weights are shown in figure 2. Here the
phase factors are such that $n = s + s^{-1} = -2 \cos 4\lambda$. The
integrable bulk weights follow in a particular limit of the
Izergin-Korepin $R$-matrix \cite{ik,n90}, with
$t_B=2 \cos \lambda$, or equivalently, with critical bulk coupling \cite{n82}
\begin{equation}
1/x^*=\sqrt{2 \pm \sqrt{2-n}}.
\end{equation}
On the other hand, the integrable boundary weights follow \cite{yb}
from appropriate
combinations of the three known reflection or $K$-matrices satisfying the
boundary version of the Yang-Baxter equation \cite{mn}.

Three inequivalent integrable sets of boundary weights are known to be
compatible with O($n$) symmetry.
One set corresponds to O-O boundary conditions \cite{bs}, with
boundary weights and equivalent critical surface couplings given by
\begin{equation}
\quad t_a = t_b = t_B \qquad \Rightarrow \qquad
y_a^* = y_b^* = x^*.
\end{equation}
Another corresponds to S-S boundary conditions \cite{by}, with
\begin{equation}
t_a = t_b =
\frac{\cos 2 \lambda}{\cos \lambda}\qquad \Rightarrow \qquad
y_a^* = y_b^* = y^*
\end{equation}
where
\begin{equation}
1/y^* = \sqrt{\pm \sqrt{2-n}}.
\end{equation}
The third set is a mixture of the above, and corresponds to
O-S boundary conditions, with
\begin{equation}
t_a = t_B, \quad
t_b = \frac{\cos 2 \lambda}{\cos \lambda}\qquad \Rightarrow \qquad
y_a^* = x^*,\quad y_b^* = y^*.
\end{equation}

The self-avoiding walk point at $n=0$ occurs at $\lambda = \pi/8$, where
$1/x^*=\sqrt{2 + \sqrt{2}}$ and $1/y^* = 2^{1/4}$ \cite{by}.
In the lattice model of the polymer adsorption transition \cite{ekb},
the self-avoiding walk has energy
\begin{equation}
E = - \epsilon L_s
\end{equation}
where $\epsilon$ is a constant and $L_s$ is the number of steps
along the adsorbing boundary, in this case the right hand side of the strip.
For the O-S boundary conditions, we thus obtain
the same critical adsorption temperature
\begin{equation}
\exp \left( \frac{\epsilon}{k T_a} \right) = y^*/x^*
= \sqrt{1+\sqrt{2}} = 1.553 \ldots
\end{equation}
as for the S-S boundary conditions \cite{by}, i.e. with adsorbing
boundaries on both sides of the strip.
Recent phenomenological renormalisation
transfer matrix calculations on the square lattice with one
adsorbing and one free boundary are consistent with this finding \cite{gb-m}.

We have previously solved the corresponding vertex model by means of
the co-ordinate Bethe ansatz for the O-O and an analytic Bethe ansatz
for S-S boundary conditions \cite{bs,yb,by}.
Proceeding in a similar manner to \cite{yb}, we find that the
eigenvalues of the
transfer matrix for the O-S boundary conditions are given by
\begin{equation}
\Lambda = \prod_{j=1}^{m}
{    {\sinh(u_j+ {\rm i}\, 3\lambda/2) \sinh(u_j-{\rm i}\, 3\lambda/2)}
                     \over
     {\sinh(u_j+{\rm i}\, \lambda/2) \sinh(u_j-{\rm i}\,\lambda/2)} }
\end{equation}
where the $u_j$ follow as roots of the Bethe Ansatz equations
\widetext
\begin{eqnarray}
\lefteqn{
 \Bigl[
  {{\sinh(u_j-{\rm i}\,\lambda/2) \sinh(u_j-{\rm i}\,3\lambda/2)}
\over{\sinh(u_j+{\rm i}\,\lambda/2)} \sinh(u_j+{\rm i}\,3\lambda/2)}
\Bigr]^N = -
  {{\sinh(2 u_j+{\rm i}\,\lambda)}
  \over{\sinh(2 u_j-{\rm i}\,\lambda)}}
} \nonumber\\
& & \times \prod_{\stackrel{k=1}{\ne j}}^{m}
{ {\sinh(u_j-u_k+{\rm i}\,\lambda) \sinh(u_j+u_k+{\rm i}\,\lambda)
   \sinh(u_j-u_k-{\rm i}\,2\lambda) \sinh(u_j+u_k-{\rm i}\,2\lambda) }
         \over
  {\sinh(u_j-u_k-{\rm i}\,\lambda) \sinh(u_j+u_k-{\rm i}\,\lambda)
   \sinh(u_j-u_k+{\rm i}\,2\lambda) \sinh(u_j+u_k+{\rm i}\,2\lambda)  }  }.
\end{eqnarray}
\narrowtext\noindent
Here $N$ is the width of the strip (e.g., $N=8$ in figure 1) and
$m$ labels the sectors of the transfer matrix, with
$m=N$ for the largest eigenvalue $\Lambda_0$. A more convenient
sector label is $\ell = N - m$.

The direct calculation of the finite-size corrections to the eigenvalue
spectrum follows a well-trodden path (see, e.g. \cite{bs} and
references therein). In this case the largest eigenvalue
in a given sector $m$ is characterised by real positive roots with
related integers $I_j =j, j = 1,\ldots,m$. Defining the
free energy per site as $f_N = N^{-1} \log \Lambda_0$, we find
\begin{equation}
f_N \simeq f_B + f_{\rm O\mbox{-}S} N^{-1} + \Delta_{\rm O\mbox{-}S} N^{-2}.
\label{cf}
\end{equation}
This result is to be compared with equation (1).
Dealing with the non-universal terms first,
\begin{equation}
f_B = \int_{-\infty}^{\infty}
             {{\sinh({1 \over 2}\pi -\lambda)x \sinh \lambda x}
        \over {x \sinh{1 \over 2}\pi x (2 \cosh \lambda x -1)}}dx
\end{equation}
is the bulk free energy \cite{bax,bs} and
\begin{eqnarray}
\lefteqn{ f_{\rm O\mbox{-}S} = {1\over 2} \log
  \Bigl(  {{1-\cos \lambda}\over{1-\cos 3\lambda}}  \Bigr)}  \nonumber\\
    &+&2 \int_{-\infty}^{\infty}
    {\sinh{1 \over 2} \lambda x
     \cosh{1 \over 4}(\pi-2\lambda) x
 \sinh{1 \over 4}(\pi-6\lambda)x
       \over
     x \sinh{1\over 2}\pi x (2 \cosh\lambda x-1) } dx
\end{eqnarray}
is the surface free energy for the mixed O-S boundary conditions.
We note however, the identity
\begin{equation}
f_{\rm O\mbox{-}S} = {\mbox{\small $\frac{1}{2}$}}
\left( f_{\rm O\mbox{-}O} + f_{\rm S\mbox{-}S} \right)
= f_{\rm O} + f_{\rm S}
\end{equation}
where $f_{\rm O\mbox{-}O}$ and $f_{\rm S\mbox{-}S}$ have been derived
previously \cite{bs,by,tbp}. The individual surface contributions are
thus
\begin{eqnarray}
\lefteqn{ f_{\rm O} = {1\over 4} \log
  \Bigl(  {{1-\cos \lambda}\over{1-\cos 3\lambda}}  \Bigr)}  \nonumber\\
    &+&2 \int_{-\infty}^{\infty}
    {\sinh{1 \over 2} \lambda x \cosh{1 \over 4} \lambda x
     \cosh{1 \over 4}(\pi-2\lambda) x
 \sinh{1 \over 4}(\pi-3\lambda)x (2 \cosh{1\over 2} \lambda x-1)
       \over
     x \sinh{1\over 2}\pi x (2 \cosh\lambda x-1) } dx
\end{eqnarray}
and
\begin{eqnarray}
\lefteqn{ f_{\rm S} = {1\over 4} \log
  \Bigl(  {{1-\cos \lambda}\over{1-\cos 3\lambda}}  \Bigr)}  \nonumber\\
    &-&2 \int_{-\infty}^{\infty}
    {\sinh{1 \over 2} \lambda x \sinh{3 \over 4} \lambda x
     \cosh{1 \over 4}(\pi-2\lambda) x
 \cosh{1 \over 4}(\pi-3\lambda) x
       \over
     x \sinh{1\over 2}\pi x (2 \cosh\lambda x-1) } dx.
\end{eqnarray}
In particular, at $n=0$ we have $\Lambda_0 = (2+\sqrt 2)^N/(1+\sqrt 2)$
with $f_B = \log (2+\sqrt 2)$, $f_{\rm O} = 0$ and
$f_{\rm S} = - \log (1+\sqrt 2)$.
Here the sign change in $f_{\rm S}$ represents an attraction towards
the adsorbing boundary.

The Casimir amplitude appearing in (1) and (\ref{cf}) is
given by
\begin{equation}
\Delta_{\rm O\mbox{-}S} = \frac{\pi \zeta \hat c}{24}
\end{equation}
where $\zeta=2/\sqrt 3$ is a lattice-dependent scale factor.
The effective central charge is
\begin{equation}
\hat c = 1 - {12 \lambda^2 \over \pi (\pi - 2 \lambda)}
\end{equation}
with $\hat c = 0$ at $n=0$.
The mixed boundary scaling index \cite{c89,bx} follows as
\begin{equation}
t_{\rm O\mbox{-}S} = \mbox{\small $\frac{1}{24}$} (c - \hat c) =
{- \pi + 8 \lambda \over 8(\pi - 2\lambda)}
\end{equation}
in agreement with the conformal invariance prediction \cite{be}.

The geometric scaling dimensions defining the surface critical behaviour
follow from the inverse correlation lengths via \cite{cx}
\begin{equation}
\xi_\ell^{-1} = \log {\Lambda_0 \over \Lambda_\ell}
\simeq {\pi \zeta X_\ell \over N}.
\label{Xf}
\end{equation}
These dimensions govern the geometric correlation
\begin{equation}
G_\ell(\mbox{\boldmath $x$}-\mbox{\boldmath $y$}) \sim
|\mbox{\boldmath $x$}-\mbox{\boldmath $y$}|^{-2 X_\ell}
\end{equation}
between $\ell$ nonintersecting self-avoiding walks tied together at their
extremities $\mbox{\boldmath $x$}$ and $\mbox{\boldmath $y$}$, which for
surface critical phenomena, are near the boundary of the half-plane\cite{ds}.
For mixed boundary conditions there is a discontinuity at the
origin between boundary conditions $a$ and $b$ corresponding
to the insertion of a boundary operator \cite{c89}.

Here the scaling dimensions $X_\ell$ are associated with the largest eigenvalue
in each sector of the transfer matrix. We find
\begin{equation}
X_\ell = {\mbox{\small $\frac{1}{4}$}}g \ell^2 +
{\mbox{\small $\frac{1}{2}$}}(g-2) \ell, \qquad
\ell = 1,2,\ldots
\label{Xos}
\end{equation}
where $\pi g = 2 \pi - 4\lambda$.
These dimensions are to be compared with
the other Bethe ansatz results. For the  ordinary (O-O) transition
\cite{bs}
\begin{equation}
X_\ell = {\mbox{\small $\frac{1}{4}$}}g \ell^2 +
{\mbox{\small $\frac{1}{2}$}}(g-1) \ell
\end{equation}
or $X_\ell = h_{\ell+1,1}$ in terms of the Kac formula \cite{c84,ds}.
On the other hand, for the special (S-S) transition \cite{by}
\begin{equation}
X_\ell = {\mbox{\small $\frac{1}{4}$}}g (\ell+1)^2 -
{\mbox{\small $\frac{3}{2}$}}(\ell+1) +
\frac{9-(g-1)^2}{4 g}
\end{equation}
or $X_\ell = h_{\ell+1,3}$ \cite{gb,d,fs}.
For the mixed O-S boundary conditions we have
$X_\ell = h_{\ell+1,1}-\mbox{\small $\frac{1}{2}$}\ell$.

At $n=0$ ($g=\mbox{\small $\frac{3}{2}$})$, corresponding to mixed adsorbing
and free boundaries in the polymer problem, (\ref{Xos}) gives
\begin{equation}
X_\ell = {\mbox{\small $\frac{3}{8}$}}\ell^2 -
{\mbox{\small $\frac{1}{4}$}}\ell.
\end{equation}
The first two values are $X_1 = \mbox{\small $\frac{1}{8}$}$ and
$X_2 = X_\epsilon = 1$.
These dimensions define critical exponents for polymers in the
upper half-plane, with one boundary condition on the positive
and the other on the negative $x$-axis, the two geometries
(the strip and the half-plane) being related
via a conformal map (see, e.g. Ref. \cite{c89}).
In particular, the number of self-avoiding walks which begin near the
origin (where the boundary conditions meet in the half-plane) scales as
$L^{\gamma_1 -1} \mu^L$ where $\mu = 1/x^*$ and the universal value
$\gamma_1 = \mbox{\small $\frac{85}{64}$}$
follows from the usual scaling relation \cite{bin,diehl}
\begin{equation}
\gamma_1 = (2 - X_1  - X_1^{\rm bulk})\, \nu
\end{equation}
where $\nu=\mbox{\small $\frac{3}{4}$}$ and
$X_1^{\rm bulk} = \mbox{\small $\frac{5}{48}$}$ \cite{n82}.
The exponent $\gamma_1 = \mbox{\small $\frac{85}{64}$}$
is to be compared with the exact values for the non-mixed cases, where
$\gamma_1 = \mbox{\small $\frac{61}{64}$}$ for a non-adsorbing
boundary (ordinary transition) and $\gamma_1 = \mbox{\small $\frac{93}{64}$}$
for an adsorbing boundary (special transition).

A detailed account of our results is currently in preparation \cite{tbp}.

\acknowledgments

It is a pleasure to thank M~N Barber, J~L Cardy and A~L Owczarek for
helpful comments.
This work has been supported by the Australian Research Council.

\newpage

\begin{figure}[htb]
\vskip 2cm
\epsfxsize = 8cm
\vbox{\vskip .8cm\hbox{\centerline{\epsffile{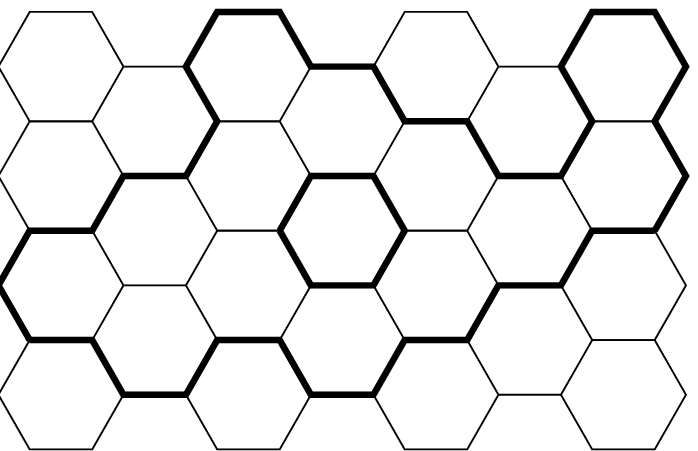}}}
\vskip .5cm \smallskip}
\caption{The open honeycomb lattice. The transfer matrix acts in the
vertical direction.}
\end{figure}
\vskip 2cm

\begin{figure}
\setlength{\unitlength}{0.00400in}%
\begingroup\makeatletter
\def\x#1#2#3#4#5#6#7\relax{\def\x{#1#2#3#4#5#6}}%
\expandafter\x\fmtname xxxxxx\relax \def\y{splain}%
\ifx\x\y   
\gdef\SetFigFont#1#2#3{%
  \ifnum #1<17\tiny\else \ifnum #1<20\small\else
  \ifnum #1<24\normalsize\else \ifnum #1<29\large\else
  \ifnum #1<34\Large\else \ifnum #1<41\LARGE\else
     \huge\fi\fi\fi\fi\fi\fi
  \csname #3\endcsname}%
\else
\gdef\SetFigFont#1#2#3{\begingroup
  \count@#1\relax \ifnum 25<\count@\count@25\fi
  \def\x{\endgroup\@setsize\SetFigFont{#2pt}}%
  \expandafter\x
    \csname \romannumeral\the\count@ pt\expandafter\endcsname
    \csname @\romannumeral\the\count@ pt\endcsname
  \csname #3\endcsname}%
\fi
\endgroup
\begin{center}
\begin{picture}(908,577)(37,149)
\thicklines
\put(129,725){\line( 3,-5){ 32.735}}
\put(161,670){\line(-3,-5){ 32.471}}
\put(164,670){\line( 1, 0){ 56}}
\put(336,324){\line(-3,-5){ 32.735}}
\put(304,269){\line( 3,-5){ 32.471}}
\put(713,216){\line( 3, 5){ 32.294}}
\put(745,270){\line(-3, 5){ 32.471}}
\put(945,536){\line(-3,-5){ 32.735}}
\put(913,481){\line( 3,-5){ 32.471}}
\put(910,481){\line(-1, 0){ 56}}
\put(224,536){\line(-3,-5){ 32.735}}
\put(192,481){\line(-1, 0){ 63}}
\put(194,478){\line( 3,-5){ 28.588}}
\put(149,480){\vector( 1, 0){ 27}}
\put(201,492){\vector( 1, 2){ 12.200}}
\put(248,480){\line( 1, 0){ 64}}
\put(312,480){\line( 3, 5){ 32.294}}
\put(314,478){\line( 3,-5){ 29.029}}
\put(333,518){\vector(-1,-2){ 12}}
\put(298,480){\vector(-1, 0){ 26}}
\put(464,426){\line(-3, 5){ 32.735}}
\put(432,481){\line(-1, 0){ 63}}
\put(434,484){\line( 3, 5){ 28.588}}
\put(389,482){\vector( 1, 0){ 27}}
\put(441,470){\vector( 1,-2){ 12.200}}
\put(488,481){\line( 1, 0){ 64}}
\put(552,481){\line( 3,-5){ 32.294}}
\put(554,483){\line( 3, 5){ 29.029}}
\put(573,443){\vector(-1, 2){ 12}}
\put(538,481){\vector(-1, 0){ 26}}
\put(705,537){\line(-3,-5){ 32.735}}
\put(673,482){\line( 3,-5){ 32.471}}
\put(670,482){\line(-1, 0){ 56}}
\put(696,445){\vector(-2, 3){ 14.923}}
\put(678,495){\vector( 2, 3){ 14.923}}
\put(823,426){\line(-3, 5){ 32.735}}
\put(791,481){\line( 3, 5){ 32.471}}
\put(788,481){\line(-1, 0){ 56}}
\put(814,518){\vector(-2,-3){ 14.923}}
\put(796,468){\vector( 2,-3){ 14.923}}
\put(251,615){\line( 3, 5){ 32.735}}
\put(283,670){\line(-3, 5){ 32.471}}
\put(286,670){\line( 1, 0){ 56}}
\put(260,707){\vector( 2,-3){ 14.923}}
\put(278,657){\vector(-2,-3){ 14.923}}
\put(369,726){\line( 3,-5){ 32.735}}
\put(401,671){\line(-3,-5){ 32.471}}
\put(404,671){\line( 1, 0){ 56}}
\put(378,634){\vector( 2, 3){ 14.923}}
\put(396,684){\vector(-2, 3){ 14.923}}
\put(586,670){\line(-1, 0){ 64}}
\put(522,670){\line(-3,-5){ 32.294}}
\put(520,672){\line(-3, 5){ 29.029}}
\put(501,632){\vector( 1, 2){ 12}}
\put(536,670){\vector( 1, 0){ 26}}
\put(610,615){\line( 3, 5){ 32.735}}
\put(642,670){\line( 1, 0){ 63}}
\put(640,673){\line(-3, 5){ 28.588}}
\put(685,671){\vector(-1, 0){ 27}}
\put(633,659){\vector(-1,-2){ 12.200}}
\put(826,669){\line(-1, 0){ 64}}
\put(762,669){\line(-3, 5){ 32.294}}
\put(760,667){\line(-3,-5){ 29.029}}
\put(741,707){\vector( 1,-2){ 12}}
\put(776,669){\vector( 1, 0){ 26}}
\put(850,725){\line( 3,-5){ 32.735}}
\put(882,670){\line( 1, 0){ 63}}
\put(880,667){\line(-3,-5){ 28.588}}
\put(925,669){\vector(-1, 0){ 27}}
\put(873,681){\vector(-1, 2){ 12.200}}
\put(592,324){\line( 3,-5){ 32.735}}
\put(624,269){\line(-3,-5){ 32.471}}
\put(601,232){\vector( 2, 3){ 14.923}}
\put(619,282){\vector(-2, 3){ 14.923}}
\put(835,215){\line( 3, 5){ 32.735}}
\put(867,270){\line(-3, 5){ 32.471}}
\put(844,307){\vector( 2,-3){ 14.923}}
\put(862,257){\vector(-2,-3){ 14.923}}
\put(458,215){\line(-3, 5){ 32.735}}
\put(426,270){\line( 3, 5){ 32.471}}
\put(449,307){\vector(-2,-3){ 14.923}}
\put(431,257){\vector( 2,-3){ 14.923}}
\put(214,326){\line(-3,-5){ 32.735}}
\put(182,271){\line( 3,-5){ 32.471}}
\put(205,234){\vector(-2, 3){ 14.923}}
\put(187,284){\vector( 2, 3){ 14.923}}
\put(177,570){\makebox(0,0){\Large $t_B$}}
\put(296,570){\makebox(0,0){\Large $s^{-1}$}}
\put(413,570){\makebox(0,0){\Large $1$}}
\put(533,570){\makebox(0,0){\Large $1$}}
\put(654,570){\makebox(0,0){\Large $1$}}
\put(774,570){\makebox(0,0){\Large $1$}}
\put(894,570){\makebox(0,0){\Large $1$}}
\put(173,370){\makebox(0,0){\Large $1$}}
\put(293,370){\makebox(0,0){\Large $1$}}
\put(413,370){\makebox(0,0){\Large $1$}}
\put(533,370){\makebox(0,0){\Large $1$}}
\put(654,370){\makebox(0,0){\Large $1$}}
\put(776,370){\makebox(0,0){\Large $s$}}
\put(898,370){\makebox(0,0){\Large $t_B$}}
\put(193,150){\makebox(0,0){\Large $1$}}
\put(317,150){\makebox(0,0){\Large $t_a$}}
\put(436,150){\makebox(0,0){\Large $s$}}
\put(847,150){\makebox(0,0){\Large $s^{-1}$}}
\put(729,150){\makebox(0,0){\Large $t_b$}}
\put(603,150){\makebox(0,0){\Large $1$}}
\put( -23,510){\makebox(0,0)[lb]{\smash{\LARGE (a)}}}
\put( -23,263){\makebox(0,0)[lb]{\smash{\LARGE (b)}}}
\end{picture}
\end{center}
\vskip 1cm

\caption{The allowed arrow configurations and corresponding
Boltzmann weights for (a) bulk and (b) surface vertices.}
\end{figure}
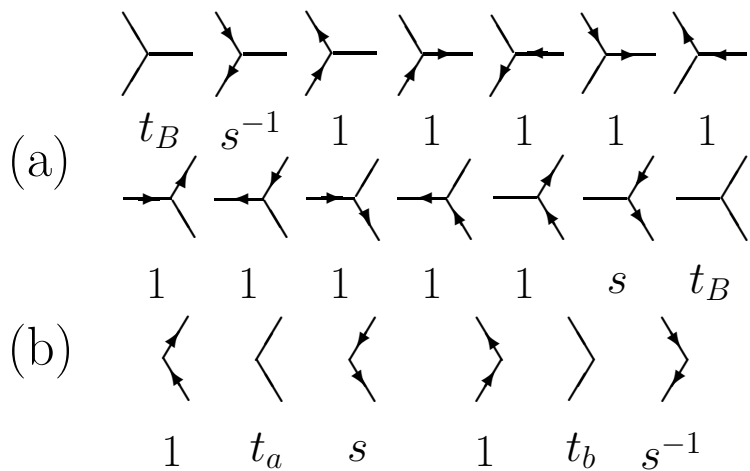

\end{document}